\documentclass{article}
\usepackage{amssymb,amsmath,amsthm}
\usepackage{graphics}
\usepackage[driver]{graphicx}

\if@twoside  m
\else
\fi \topmargin 5mm \headheight 0mm \headsep 0mm \textheight
225truemm \textwidth 150truemm
\newcommand{\cf}{{\em cf.}}

\begin{document}

\title{Force Plate Monitoring of Human Hemodynamics}
\author{J.~K\v{r}\'\i\v{z}$^1$ and P.~\v Seba$^{1,2}$
\\
\it $^{1}$University of Hradec Kr\'alov\'{e}, Rokitansk\'eho 62 \\
\it 500 03 Hradec Kr\'alov\'{e}, Czech Republic\\
\it $^{2}$Institute of Physics, Academy of Sciences of the Czech
Republic\\  \it Cukrovarnick\'{a} 10, CZ - 162 53 Praha, Czech
Republic}
\date{}
\maketitle
\begin{abstract}
We show that the recoils of the body caused by cardiac motion and
blood circulation provide a noninvasive method capable to display
the motions of the heart muscle and the propagation of the pulse
wave along aorta and its branches. The results are compared with the
data obtained invasively during a heart catheterization. We show
that the described noninvasive method is able to determine the
moment of a particular heart movement or the time when the pulse
wave reaches certain morphological structure.
\end{abstract}

\section{Introduction}

The objective of this work is to demonstrate that the force changes
evoked by the cardiac activity and measured by the force plate can
be used to trace the motion of the heart muscle and the pulse
propagation along the aorta. In particular the recoils caused by the
pulse wave scattering on certain morphological structures (for
instance on the aortal arch) can be recognized and used to determine
the time points when a pulse wave reaches given location.

The idea itself is not new. The usage of body micro movements for
extracting information about the cardiac activity is the principle
of ballistocardiography - a field that is worked out for more then
50 years \cite{trefny}.  Nowadays the body motion is usually traced
by a sensitive low frequency accelerometers fastened on the sternum
\cite{McKay}. In the past various method has been used ranging from
pendulous bed to induction coils. (Movements of the body occurring
synchronously with the pulse were first reported already by Gordon
(1887) using an ordinary spring weighing machine \cite{G}). Similar
methodology derived from the field of seismology is known as
seismocardiography. \cite{IKK}.

To elucidate the cardiac dynamics more closely a three dimensional
ballistocardiography was developed at late eighties \cite{Soames}.
Using a special platform the body motion was traced in three
mutually orthogonal directions.  More recently similar measurements
have been done also in sustained microgravity during cosmic flights
- see for instance \cite{prisk}. A multidimensional measurement
allows a better classification of the vascular recoil since  the
arteries form a three dimensional system leading to micro movements
in different directions.

\section{Method}

The standard ballistocardiography is established on registering the
\it motion \rm (or acceleration) of the body. We use another
technique based on a direct measurements of \it forces and moments
\rm induced by the cardiac activity. A similar method was described
in \cite{berme}. To measure the forces we used a stiff bed mounted
on the top of a standard Bertec force plate, model~4060A, equipped
with strain gage transducers.  The force plate is capable to detect
independently the force and  moment components in the $x$, $y$, and
$z$ directions with a precision of 0.1 N and Nm respectively. A
volunteer was reclining on the bed on the back without voluntary
movements. So the recorded force and moment changes were evoked
mainly by the mass motion inside his body. (Due to the relaxed
position the body tremor can be neglected.) Simultaneously with the
force changes the ECG signal registered by a one-channel
electrocardiograph was recorded. All signals were digitized with a
AD converter with sampling rate of 1 kHz. This yields together to a
seven dimensional time series that has to be inspected.

The clue is however not contained in the measured data but rather in
the method we used to uncover the underlying processes. We
understand the time series measured by the force plate as coordinate
projections of a six dimensional geometric object - a signal curve.
To extract the information contained in the signal curve we will
investigate its geometric properties. In particular we focus on the
geometric invariants.

Geometric invariants do not change when the curve is rotated,
translated or its parametrization is changed. This is in contrast
with its coordinate projections (measured signals) that change under
such transformations. It is clear that if we, for instance, slightly
change the position of the measured body, the intrinsic
hemodynamical process will be not affected. But  the individual
force components measured by the plate will change. The geometric
invariants of the signal curve will, however, remain unchanged.

In a three dimensional space geometric invariants are usually used
as shape and object descriptors and apply, for instance, in
computerized object recognition -see \cite{recog}. The idea behind
this effort is simple: the invariants set the clue that remains
unchanged when the object moves, rotates or the view perspective
changes. Moreover they help to distinguish different object shapes
etc. These invariants usually describe the vertices and edges of the
object. In fact there is  an impressive amount of various invariants
used in the three dimensional space. But not all of them suitable in
higher dimensions.

Here we will use invariants known as Cartan curvatures
\cite{gallier} that apply naturally in spaces with arbitrary
dimension. Mathematically this invariants are  based on the concept
of local coordinate frames  associated with a curve. In a given
point one axis of the local frame is tangential to the curve, the
second axis represents the normal, the third is the binormal etc.
Roughly speaking when we consider a smooth $n$-dimensional
curve~$c(t)$, there is for each parameter $t$ an orthogonal frame
$(E_1(t),\ldots,E_n(t))$ of $n$-dimensional vectors such that the
$k$th derivative $c^{(k)}(t)$ of the curve $c(t)$  can be expressed
as  a linear combination of its first $k$ vectors $(E_1,\ldots,E_k)$
, $1 \leq k \leq n-1$. The family $(E_1(t),\ldots,E_n(t))$ is called
the distinguished Frenet frame. The curvatures $\kappa_i$, $i =
1,\ldots n-1$ of the curve~$c$ are defined by Frenet-Serret formulae
as
\begin{equation*}
\kappa_i(t) = \frac{E_i'(t) \cdot E_{i+1}(t)}{\|c'(t)\|}\,.
\end{equation*}
So the curvatures characterize local changes of the coordinate
system related with the curve. In our case the signal curve is
6~dimensional and is characterized by 5~curvatures.

Mechanical events like a heart beat or a scattering of the pulse
wave on an arterial bifurcation lead to recoil and is registered by
the force plate. It might be difficult to trace up this response
just by inspecting  the  time series representing the individual
force and moment components. But the events  change the geometry of
the total signal curve. Hence they will be visible as a changeover
of its invariants. We use this strategy to trace up the footprints
of the hemodynamics in the force plate signal. It turns out that for
this purpose it is enough to investigate the first Cartan curvature
$\kappa_1(t)$ only.

\section{Results}
We have investigated a sample of 20 healthy young adult males. They
were asked to recline on the back in a relaxed position without
voluntary motion.  The measurement with an 8 minute data acquisition
started always after a calm down period of at least 5 minutes

The cardiac cycle is triggered by the electric activity represented
in the ECG signal. We used (in a consent with the literature
\cf~\cite{Sramek}) the absolute maximum of the ECG R wave as the
starting point. The response of the plate was registered for time
intervals that started 300 ms before a particular R wave and
terminated 600 ms after it. This time interval covers usually the
whole cardiac cycle. During the time period the heart contracts and
the blood ejection takes place. The corresponding pulse wave shoots
along the aorta passing its main branchings. Finally the heart
muscle relaxes in the diastolic part of the cycle.

In all measured cases we discovered several clear changes (maxima)
of the curvature that come up with constant time delays with respect
to the R wave. It is natural to assume that such are linked to
events related to cardiovascular system. This expectation is
strengthened by the fact that the obtained time delays are similar
for all measured subjects. In addition there is a clear correlation
between the observed event (change in the curvature) and the size of
the stature of the measured person: longer persons embody longer
time delays. This fits well with the hypothesis that relates the
changes of the geometry of the signal curve to the motion of the
heart muscle and to the scattering of the ejected blood on various
morphological structures of the arterial system.  We get naturally
longer time delays between the R wave and the specific scattering
event for taller persons. Simply because the distance to be passed
by the pulse is longer.

To get a direct evidence that supports the above hypothesis is,
however, not easy. It is rather simple to observe the motion of the
heart muscle (for instance with echography) and compare the results
with the signal changes of the force plate. But a noninvasive
observation of the pulse wave propagation is difficult. The process
is very quick and the time resolution of a standard echograph fails
to resolve it- see \cite{pulse wave velocity with ECHO}, \cite{pulse
wave velocity with ECHO2}. It can be noninvasively examined only by
special-purpose devices that were inaccessible for us.

To find the way out we asked for collaboration physicians and
patients from the catheterization unit of the University hospital in
Hradec Kralove. Before the examination  on the catheterization unit
we inspected the patient on the force plate using the method
described above. Later on the same patient was examined with a
catheter equipped with a pressure transducer. During the
investigation the blood pressure inside various locations of the
aorta and the left ventricle was recorded. Since the ECG signal is
recorded simultaneously with the pressure it is possible to set the
time lag between the R wave and the arrival of the pulse wave at a
given location.

The arrival of the pulse wave is characterized by an rapid increase
of the local pressure. The pulse shape is however not rectangular.
We define the time of the pulse arrival as the point of the maximal
pressure increase i.e.  as the maximum of the derivative of the
measured pressure signal. Similar method was used in \cite{micro}
where pressures were recorded at five regions from the ascending
aorta to the iliac artery.

The knowledge about the pulse wave propagation obtained invasively
during the catheterization  enables us to verify the hypothesis that
relates the observed sudden changes of the curvature of signal curve
 to the internal hemodynamical process. To do this we compared the
arrival times of the pulse wave measured during the catheterization
with times characterizing the localizations of the curvature peaks
by the same person.

We compared the curvature peaks with the pulse arrival times  on the
following arterial structures:

\begin{itemize}
\item truncus brachiocephalicus
\item top of the aortal arch
\item the place where the aorta comes through diaphragm
\item splitting of arteria renalis
\item the bifurcation of the abdominal aorta to
the common iliac arteries
\item the bifurcation of the common
iliac arteries into the artheria iliaca interna and artheria
iliaca externa
\item junction of the artheria iliaca externa and
artheria femoralis
\end{itemize}

\begin{figure}[h]
\vspace{0.5cm} \hspace{-2.75cm}
\begin{center}
\includegraphics[angle=0,width=420pt,height=271pt]{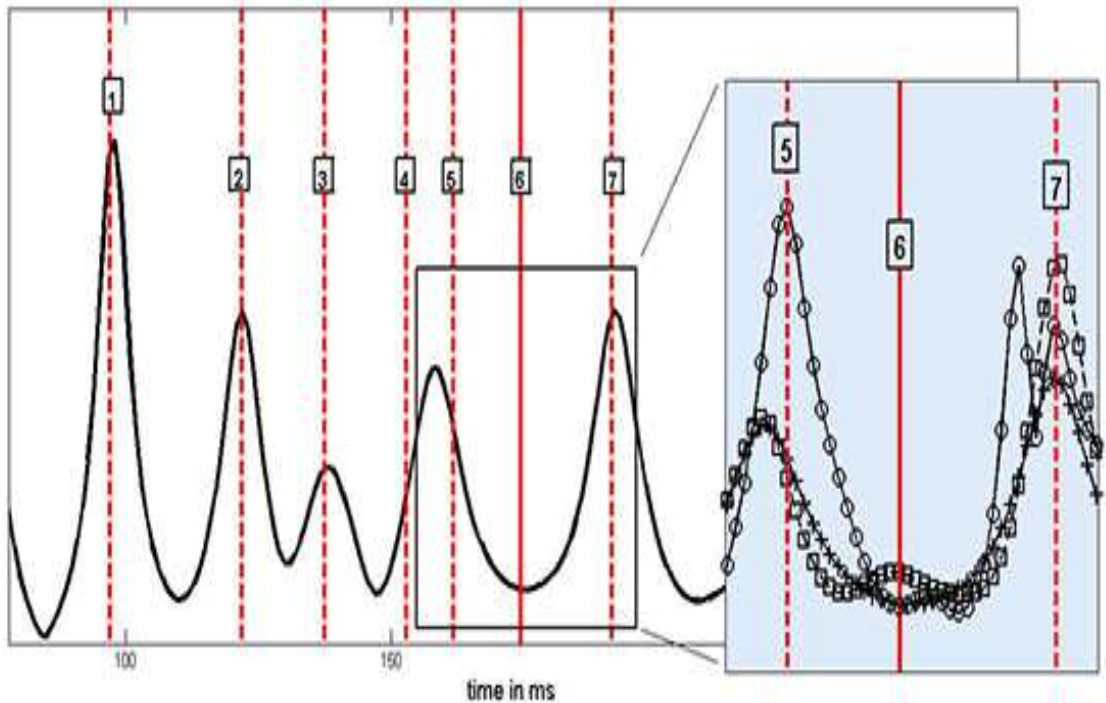}
\end{center}
\caption{The first curvature of the force plate signal is plotted
and compared with times when the pulse wave reaches certain
morphological structures (labeled vertical lines). The labels stay
for: 1 - aortal root; 2 - truncus brachiocephalicus; 3- top of the
aortal arch; 4 - below the diaphragm; 5 - splitting to renal
arteries; 6 - the bifurcation of the abdominal aorta to the common
iliac arteries; 7 - junction of the artheria iliaca externa and
artheria femoralis. The time counting starts at the maximum of the R
wave. The insert displays the first (crosses), second (squares) and
third (circles) curvatures referring to a marked time interval.}
\label{hemo}
\end{figure}

The results are encouraging. Some of the pronounced peaks observed
in the curvatures appear exactly at times when the pulse wave reach
anatomically distinguished positions and is scattered.  Typical
results are plotted on figure~\ref{hemo}.

The agreement is remarkable. Especially for isolated events. When
the recoil is weak it is usually not resolved by the first
curvature. A typical example is the arrival of the pulse wave to the
aortal iliac bifurcation where the aorta is divided into the two
common iliac arteries. The pulse wave transmission on this structure
is very smooth and the related recoil is small. So the process is
usually not visible with the help of the first Cartan curvature. But
it is uncovered when a second curvature is used - see the insert of
the figure ~\ref{hemo}.  Another problematic situation occurs when
there are several subsequent scatterings of the pulse wave within a
very short time interval. This happens below the diaphragm. Here the
pulse wave scatters on the diaphragm, on the celiac artery splitting
, mesenteric artery and on the renal arteries. All there structures
are localized close to each other.  The first curvature does not
resolve the events and joints them into one peak and higher Cartan
curvatures have to be used. The insert of the figure ~\ref{hemo}
shows that the third curvature resolves correctly the pulse
scattering on renal arteries.

There are, however,  curvature peaks that appear before the QRS
complex and cannot be related to the pulse wave propagation. Other
curvature peaks, on the contrary, refer to time delays that are
longer than the time during which the pulse wave propagates along
the arterial tree. We will demonstrate that these peaks are directly
associated with the motion of the heart muscle.

One of the  signals obtained during the catheterization is the left
ventricular pressure. Since the heart motion leads to well defined
pressure changes this signal can be used to trace the mechanical
myocardial motion.

The heart motion starts with the contraction of the left atrium
which raises the left atrium pressure and leads finally to the
ejection of the blood from the left atrium into the left ventricle
(the A wave). This process is measurable as a small pressure
increase in the left ventricle. The atrium contraction is followed
by the contraction of the ventricle. The ventricular pressure
increases quickly and eventually exceeds the atrium pressure. This
leads to a back flow that closes the mitral valve. At this stage
both the mitral and aortic valves are closed. The continuing rapid
ventricular contraction (isovolumic contraction) leads to an
increase of the  pressure that eventually opens the aortic valve and
ejects the blood into the aortic root. Similar process (with a
possible small time delay) takes place also in the right part of the
heart. However the pressure changes are much smaller there and
result to minor force response of the plate.

There are usually 2 curvature peaks preceding the Q wave. These
peaks appear after the P wave which marks the electric trigger for
the atrial contraction. There is a clear link between the peaks and
small pressure changes in the left ventricle. The first peak is
related with a small increase of the left ventricular pressure
caused by the atrial contraction. The position of this peak
coincides with the local maximum of the derivative of the pressure
signal measured in the left ventricle between the P and Q waves -
see the inset $a)$ in the Figure 2.  The second curvature peak is
associated with a small plateau-like local maximum of the left
ventricular pressure and follows closely  the Q wave of the ECG
signal. We assume that this peak reflects the force changes
associated with the beginning of the ventricular systole.

The R wave is followed by a sequence of 2-3 curvature peaks. The
most important is the first of them. It marks the onset of the rapid
pressure increase in the left ventricle (isovolumic contraction) and
reflects the enclosure of the bicuspid valve (when a heart sound is
measured this peak coincides exactly with the beginning of the first
heart sound). The remaining  peaks of this series  are associated
with heart muscle motion due to the isovolumic contraction. It  ends
with the aortal valve opening. The valve opening is,however, not
instantaneous. It starts at the moment when the ventricular pressure
equals to the pressure inside the aortal root. But, as the systole
holds on, the ventricular pressure may - for a short time period -
exceed the aortal pressure. Whenever the aortal valve opens fully
these pressures match. The opening of the aortal valve  is clearly
distinguished as an anacrotic notch  - a bump in the measured aortal
pressure. At the same time the curvature of the force plate signal
curve shows a sharp and high peak. It stays for the aortal valve
opening and for the beginning of the blood election into the aortal
root.

The discussed  systolic part of the cardiac cycle is characterized
by a rapid heart muscle contractions and pulse wave propagation. The
diastolic part of the cycle, during which the heart muscle relaxes
and is passively filled with blood, does not contain any abrupt
motions. Nevertheless it is also related with an internal mass
motion and hence  measurable by the force plate.

From the point of view of the measured pressure the ventricular
diastole is characterized by two main event. It starts with the
closure of the aortal valve which is recognized as the dicrotic
notch in the aortal pressure signal. During the first part of
diastole the ventricular pressure falls down rapidly. This pressure
drop stops at the moment when the mitral valve opens and the blood
flows into the ventricle - passive ventricular filling. The atrial
diastole precedes the ventricular one and takes place already during
the ventricular systole. So it is superimposed by it.

The curvature of the force plate signal detects the ventricular
diastole as a sequence of several overlapping peaks. This sequence
starts with the aortal valve closure that is marked by the dicrotic
notch and/or (if the sound is recorded) by the second heart sound.
It ends short after the termination of the sudden drop of the
ventricular pressure, i.e. after the opening of the mitral valve.

This sequence of peaks is followed by one or a couple of well
pronounced curvature maxima representing the first (rapid) phase of
ventricular filling i.e. the phase when the blood accumulation in
the atrium during the atrial diastole flows rapidly into the
ventricle. This phase is followed by the atrial systole that has
been already described.

\begin{figure}[h]
\vspace{0.5cm} \hspace{-2.5cm}
\begin{center}
\includegraphics[angle=0,width=420pt,height=261pt]{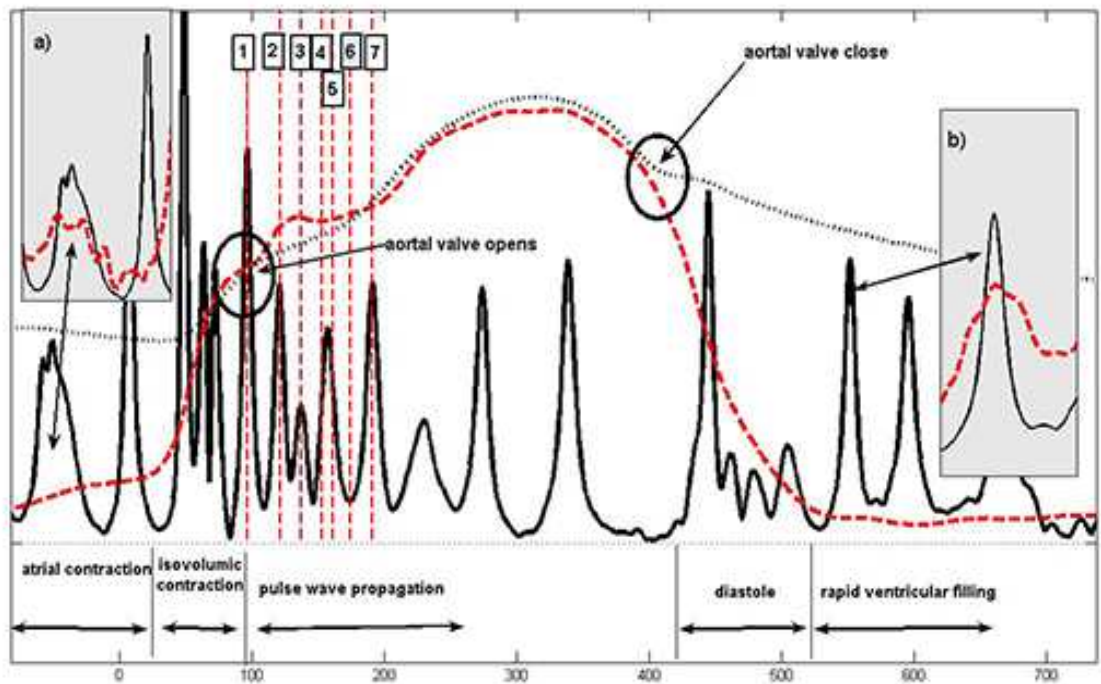}
\end{center}
\caption{The first curvature is plotted as a function of time (full
line)  together with the pressure measured in the left ventricle
(dashed line) and in the aortal root (dotted line). The moments when
the pulse wave reaches certain morphological structures are labeled
with dashed vertical lines. The numbers have the same meaning as on
the Figure 1. The insert a) display the marked peak (atrial
contraction) together with the derivative of the ventricular
pressure. The insert b) shows the curvature peak corresponding to
the rapid filling together with the derivative of the ventricular
pressure.} \label{trnka2}
\end{figure}

An typical example of the obtained curvature and pressures measured
in the left ventricle and aortal root is plotted on the
figure~\ref{trnka2}.

\section{Discussion}
\begin{figure}[h]
\vspace{0.5cm} \hspace{-2.5cm}
\begin{center}
\includegraphics[angle=0,width=420pt,height=185pt]{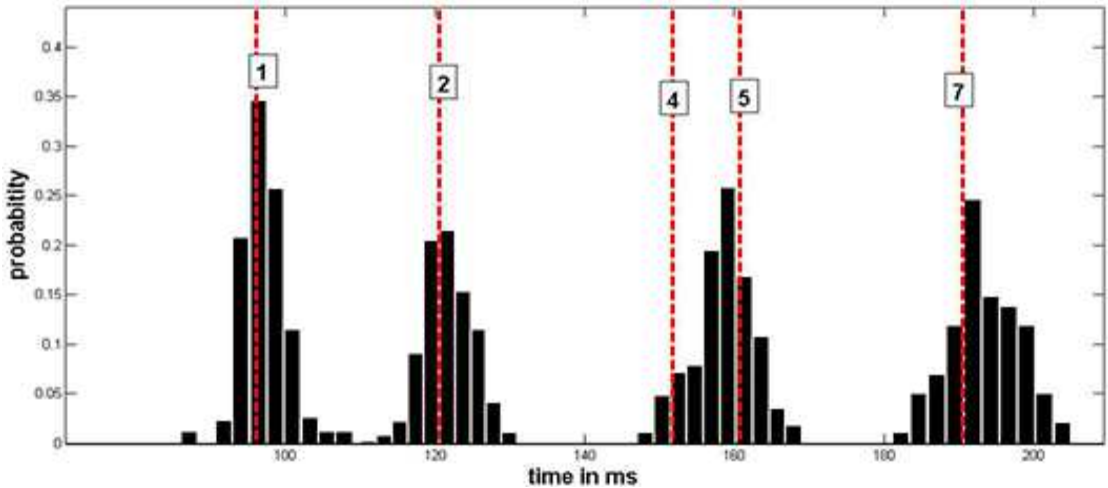}
\end{center}
\caption{The probability distributions of the peak localizations of
the curvature obtained during 300 subsequent heart beats. The
distributions refer to the peaks 1,2,4,5 and 7 of the
figure~\ref{hemo}. The times of the pulse wave arrival obtained
during catheterization are plotted as dashed lines} \label{distrib}
\end{figure}

The curvature pattern described in the previous section is
reproducible. Herewith we mean the following:
\begin{itemize}
\item It is stable during the heart cycle. This means that the
curvature peaks appear with almost constant time delay with respect
to the R wave of the ECG signal. The probability distributions of
time delays referring to the peaks 1,2,4,5 and 7 of the
figure~\ref{hemo} are plotted on  the figure~\ref{distrib}.

\item As  mentioned repeatedly in this paper:  the curvatures are geometric
invariants of the signal curve. So it is not surprising that the
position of the peaks is not sensitive to the particular position of
the body as long as the internal hemodynamical process is not
changed. We measured the subjects on the back and on the belly. The
position of the peaks is the same in both cases. But their hight
might change.

\item In several cases we measured the force plate response once
more after a time interval of  three months. The position of the
curvature peaks was stable, i.e. it has not changed when the same
patient was measured again.

\item The geometric properties of the signal curve are analogous
for different people. For instance in a sample of 25 examined
volunteers the curvature peak that stands for the blood injection
into the aortal root appeared in the mean 80 ms after the R wave
with a standard deviation of 8 ms.

\end{itemize}

All this together shows that the changes of the  geometric
invariants of the signal curve obtained by a force plate are
suitable to display the internal hemodynamical processes. It has to
be stressed that the method is fully noninvasive. All what is needed
is that the patient reclines quietly on a bed equipped with force
transducers. Then the pulse wave propagation and the mechanical
strokes of the heart muscle can be monitored in real time (online).
This may be used to trace out for instance the immediate influence
of an administered drug etc.

We used the first Cartan curvature to reveal the information
contained in the signal curve. Similar results are obtained also
when higher curvatures are inspected. The main difference is that
higher curvatures are more sensitive to the signal noise. On the
other hand - as demonstrated above - some smooth processes are
better visible with the help of higher curvatures.

Similar effect can be reached when the body position is changed.
When the subject is measured on the belly instead on the back some
recoils are intensified and the related curvature maxima are more
pronounced. A typical example is the celiac artery splitting. It is
badly visible when the subject reclines on the back. But it leads to
well pronounced peak when the measurement is done on the belly.

\section{Acknowledgements}
This work was supported by the project LC06002 of the Ministery of
Education, Youth and Sports of the Czech Republic, by the GACR grant
202/06/P130 and by the Theoretical physics supporting fund in
Slemeno. The measurement were done on the force plate belonging to
the Department of Rehabilitation of the Faculty of Medicine of the
Charles University in Hradec Kralove. The authors are  grateful to
Jan Skala and Petrof company for the help with the
phonocardiographic measurements. Cooperation with the heart
catheterization unit of the University hospital in Hradec Kralove is
also gratefully acknowledged.

\end{document}